# I. Introduction

Alfalfa (*Medicago sativa* L.) is a perennial plant which belongs to the Fabaceae family. It is known for its high protein content, which makes it a valuable source in plant-based diets (Hadidi *et al.*, 2023). Research supports its potential in food applications and human nutrition because alfalfa leaves contain high levels of vitamins and minerals (Almuhayawi *et al.*, 2021; Tucak *et al.*, 2021). In particular, the young seedlings, commonly referred to as sprouts, are prized for their exceptional nutritional profile, which develops during the germination of seeds. Alfalfa sprouts are widely consumed, either in raw form in salads or as lightly cooked (Rajkowski *et al.*, 2001). However, various factors can inhibit the germination of alfalfa seeds such as impermeable seed coat, dormancy, microbial contamination, and environmental contrast (Kameswara *et al.*, 2017). These challenges not only delay the development of sprouts but also increase production costs. Therefore, enhancing seed germination is one of the most efficient strategies to improve the establishment and growth of alfalfa sprouts.

Seed priming is a commonly employed technique in commercial agriculture to improve seed vigor, germination potential, and stress tolerance (Paparella *et al.*, 2015). This is in spite of a wide range of physical, chemical, and biological treatments that are available. The chemical-based priming methods are increasingly restricted because of their potential adverse effects on seed storage and the associated risk of environmental pollution (Bera *et al.*, 2022). In response, physical priming techniques have gained attention as eco-friendly and cost-effective alternatives for enhancing seed germination and seedling growth. Techniques such as cold plasma (CP) treatment and ultraviolet (UV) radiation offer significant advantages by not only improving seed performance but also by contributing to sustainable agricultural practices. These methods represent a new frontier in green technology, providing a safer and more efficient approach compared to conventional chemical treatments.

CP, also known as nonthermal plasma, is a partially ionized gas that contains a mixture of electrons, ions, atoms, and reactive species of short lifespan (i.e. lower than 1 ms such as OH and NO radicals) and longer (i.e. ozone and nitrogen dioxide) (Bourke *et al.*, 2018). Unlike hot plasma, where all particles are in thermal equilibrium and at very high temperatures (e.g. sun, lightnings), CP features electrons that are highly energetic, while the heavier ions and neutral particles remain at or near room temperature. This means that CP can be generated and used at relatively low temperatures, making it suitable for various applications such as medicine, food decontamination, and agriculture without causing thermal damage to surrounding materials or tissues. In the case of agriculture application, CP is considered an economic, rapid, and environmentally friendly method to stimulate seed germination and seedling growth. It has been successfully applied to improve the parameters of seeds related to germination like wettability, permeability, and water absorption (Benabderrahim *et al.*, 2024; Li *et al.*, 2017).

UV irradiation, also referred to as photostimulation, is known as a green technology that activates biological parameters of seeds (Kondrateva *et al.*, 2019). In addition to the photosynthetic active radiation (400–700 nm), plants are exposed to three types of UV radiation: UV-A (320–390 nm), UV-B (280–320 nm), and UV-C (below 280 nm) (Foroughbakhch *et al.*, 2019). Extensive research has been conducted to study the biological impacts and mechanisms of UV radiation (Semenov *et al.*, 2020). While UV-C radiation is mostly filtered by the ozone layer in the stratosphere, strong doses of UVA, UV-B, or UV-C can negatively impact plant health. In contrast, low levels of UV radiation, particularly in seed treatments, offer a sustainable and safe method to enhance plant productivity and stress tolerance (Hernandez-Aguilar *et al.*, 2021).

Foroughbakhch *et al.* (2019) have reported that sensitivity of seeds to UV-B and UV-C radiation varies across species during germination tests. In agriculture, the nonionizing UV-C radiation has demonstrated positive effects on seed health, germination, and seedling vigor (Neelamegam and Sutha, 2015; Sukthavornthum *et al.*, 2018). Furthermore, some studies have revealed that UV-B radiation, when carefully applied, can significantly influence plant development, biomass production, and metabolism (Ozel et al., 2021). Although fewer studies have examined plant responses to UV-A, there is evidence that it could influence the biomass production of plants, with varied responses observed between plant organs (Verdaguer et al., 2017). Overall, detailed research on UV-A effects on seed irradiation remains limited compared to UV-B and UV-C.

In this context, the present work aims to study the effects of CP and UV irradiation treatments (UV-A, UV-B, and UV-C) on alfalfa seed germination, focusing on key parameters such as germination rate, imbibition, electrolyte leakage (EL), and wettability. In addition, the study examines seedling growth (epicotyl and hypocotyl development) and assesses the phytochemical composition of the sprouts, including chlorophyll (Chl) and carotenoid contents, total soluble sugar, starch, proteins, polyphenols, and flavonoids. By exploring these eco-friendly technologies, this study seeks to advance the understanding of how to improve sprout production and contribute to sustainable practices in the agricultural industry.

# II. Material and Methods

### II.1. Seed material

Alfalfa seeds (*Medicago sativa L.*), from a local cultivar, were harvested from the experimental field of the Arid Regions Institute of Medenine, Tunisia during the summer of 2023. Only the vigorous seeds were chosen. After harvesting, the seeds were stored at −4°C until the commencement of plasma and UV treatments.

### II.2. Seed treatments with cold plasma (CP) and UV radiation

CP of ambient air was generated in a dielectric barrier device (DBD) operating in a plan-to-plan configuration. The system consisted of a 2 mm thick dielectric barrier and two electrodes: a high-voltage stainless-steel mesh electrode (30 × 40 cm²) and a grounded counter-electrode made of bulk alumina. The alfalfa seeds were placed in the interelectrode region characterized by a gap of 1 mm. The mesh electrode was powered by a high-voltage generator comprising a function generator (ELC Annecy France, GF467AF) and a power amplifier (Crest Audio, 5500W, CC5500), applying a sine-wave voltage of 7 kV at 500 Hz. The alfalfa seeds were exposed to this treatment for 5 minutes.

For UV radiation treatments, seeds were exposed to a 3UV-36 lamp emitting UV-B (310 nm) and UV-C (254 nm) radiations with intensity of 0.06 W m-2 and 0.03 W m-2, respectively (Díaz-Leyva et al., 2017). UV-A (370 nm) radiation was applied at an intensity of 2 W m-2 (Lim et al., 2021). Each UV irradiation treatment was carried out at 25°C for 15 minutes.

### II.3. Electrolyte leakage and water uptake (WU) of seeds

To evaluate the membrane integrity of untreated and treated seeds, EL measurements were carried out, based on the method described by Lazar et al. (2014). For this, three replicates of 0.1 g of seeds were placed into 10 mL of deionized water and incubated in a water bath at 25°C. The conductivity of the solution was determined using a WTW Cond 730 conductivity meter (GeoTech, Denver, CO, USA) periodically after 24 hours, 48 hours, and 72 hours. Results are expressed in mS/cm.g of seed and represent the mean of three replicates ± standard error (SE).

The WU was determined by placing 0.1g of healthy seeds (initial dry weight, Wi) in 10 mL of distilled water (pH = 7) for 24 hours, which corresponds to the estimated time for 50% germination. After 24 hours, the seeds were removed, surface-dried using filter paper, and their fresh weight (WS) was recorded. The WU was calculated according to the formula 1:

$$WU = \frac{W_S - W_i}{W_S} \times 100 \qquad (1)$$

### II.4. Degree of wettability (WCA) of seeds

The wettability of the outermost layers of alfalfa seeds was evaluated using drop shape analysis. The sessile drop technique was employed, where a single droplet (1.5 µL, distilled water) was carefully applied to the surface of each seed. Then, the ImageJ program (https://imagej.nih.gov/ij/plugins/contact-angle.html) was used to calculate the water contact angles (WCA).

### II.5. Germination kinetics

The germination tests were carried out under controlled conditions at 25°C in darkness. Fifty seeds from each treatment group (C, CP, UV-A, UV-B, and UV-C) were placed on a petri dish lined with a single layer of filter paper (day 0). The petri dishes were moistened with 5 mL of distilled water every 2 days throughout the experiment. Each treatment was tested in triplicate across three separate experiments. Following germination, the sprouts were grown in regulated environment conditions in a culture room (40% humidity, 25°C temperature, and a 16–8 h light–dark cycle).

The number of germinated seeds per petri dish was counted every 7 hours. Seeds were considered to have germinated when the radicle length reached or exceeded 2 mm (Manmathan et al., 2013). The germination experiment was performed in three replicates (a total of 150 seeds per treatment). The germination rate ($G_\%$) was calculated according to the following formula 2:

$$G_\% = \frac{100}{N} \times \begin{bmatrix} Number\ of\ seeds \\ germinated\ in\ x\ hour \end{bmatrix} \qquad (2)$$

## II.6. Seedling growth and sprout biomass

Six days after sowing, the root length (RL) and hypocotyl length (HL) of 10 randomly selected germinated seeds (from each treatment group) were photographed and measured with ImageJ software. Sprout biomass was assessed through two parameters: fresh weight (FW) and dry weight. The sprout FW (SFW, mg/seedling) was determined by weighting 25–30 seedlings, and the total weight was divided by the number of seedlings. To assess the dry weight, the seedlings were dried in an oven set at 80°C overnight and then weighed to determine the sprout dry weight (SDW, in mg/seedling).

## II.7. Chls and carotenoids contents in sprouts

The pigment contents were determined using fresh matter. For this, 100 mg of seedling aerial parts were cut into small discs and immersed in 80% acetone (5 mL). The mixture was kept at 4°C for 72 hours. Then, the extract was filtered, and the absorbance was measured with a Beckman spectrophotometer at 470 nm, 646 nm, and 663 nm. The Chls and carotenoids contents are expressed as mg/g FW (Ghazouani *et al.*, 2021).

## II.8. Protein contents in sprouts

Bradford's (1976) technique was used to determine the protein content. To prepare the color reagent, 100 mg of Coomassie Brilliant Blue G-250 (Sigma-Aldrich Co.) was dissolved in 50 mL of 95% ethanol, followed by the addition of 100 mL of 85% phosphoric acid. After being diluted and filtered, the resulting solution was used as the color reagent for protein measurement. Standard solutions of reagent grade BSA (Equitech-Bio, Inc., Kerriville, TX) were prepared, ranging from 0 to 400 µg of protein. Both the samples and standards were incubated for 5 minutes, covered with parafilm, to allow for color development. The absorbance of the samples and standards was measured at 595 nm using a spectrophotometer (Anthelie Advanced, Microbeam, S.A.). To determine the protein contents of the samples, a least squares regression analysis was performed to correlate the standard protein concentrations with their absorbance measurements.

## II.9. Total soluble sugar and starch contents in sprouts

Soluble sugar and starch content were determined according to the anthrone method (Staub, 1963). In order to extract sugars, 10 mg of fresh sprouts were mixed in 1 mL of 80% ethanol and incubated in a water bath at 70°C for 30 minutes. After cooling, the solution was centrifuged at 9000 tr/minute for 15 minutes at 4°C. This process was repeated three times to ensure thorough extraction. The supernatants and the pellets were recuperated and used to estimate the contents of sugar and starch, respectively.

For the determination of sugar content, the anthrone reagent was prepared by dissolving 2 g of anthrone in 100 mL of concentrated sulfuric acid (36 N). Then, 2 mL of this anthrone reagent was added to 1 mL of the supernatant. The mixture was stirred and incubated in a water bath at 100°C for 10 minutes. Immediately after incubation, the solution was cooled on ice to halt the reaction. The absorbance of the mixture was measured at 640 nm.

For the determination of starch content, the retained pellet was mixed with 1 mL of 80% ethanol and centrifuged twice at 9,000 rpm for 10 minutes at 4°C. The starch extraction (1 mL of supernatant) was performed using perchloric acid (30 µL). The mixture was kept in ice for 30 min. Subsequently, 2.5 mL of anthrone was added to 1.2 mL of 80% ethanol and 0.1 mL of the extract. After agitation, the solution was incubated at 100°C for 10 minutes. After cooling, the absorbance was recorded at 640 nm. A calibration curve was prepared using a glucose solution as standard, and a control was prepared under the same experimental conditions using 80% ethanol. The sugar and the starch contents are expressed as mg per gram of fresh matter (mg/g FM).

## II.10. Total phenolic and flavonoid contents in sprouts

The extraction of phenolics and flavonoids was carried out by maceration at 25°C for 24 hours using 1 g of dry matter in 10 mL of ethanol 60%. Then, the mixture was filtered, and the obtained extract was kept at 4°C until analysis.

The total phenolic content (TPC) of the ethanolic extracts from sprouts was determined using the Folin–Ciocalteu reagent method. For the calibration standard, concentrations of gallic acid ranging from 0 to 600 µg/ mL were employed. The results are expressed as milligram gallic acid equivalent (GAE) per 100 g of sprout dry weight (mg GAE/100 g DW).

The total flavonoids content (TFC) was determined using a colorimetric method based on the formation of a flavonoid–aluminum complex, with maximum absorbance at 430 nm. Each sample extract (1 mL) was mixed with 1 mL of sodium nitrite ($NaNO_2$) 0.5 M and 150 µL $AlCl_3$ 0.3 M. Then, the absorbance was measured at 430 nm after 15 minutes of incubation at room temperature. Rutin was used as the standard to generate the calibration curve, and the total flavonoid content was expressed as milligrams of rutin equivalents per 100 grams of dry weight (mg RE/100 g DW) (Djeridane *et al.*, 2006).

## II.11. Statistical analysis

The XLSTAT software v.2019 (https://www.xlstat.com) was used to perform the statistical analysis using one-way ANOVA. The results were presented as means ± standard deviation (± SD). To estimate significant differences between treatments, Duncan's multiple range test ($p < 0.05$) was performed. All measurements were achieved at least in triplicate.

# III. Results and Discussion

## III.1. Kinetics of alfalfa seed germination

The effects of CP and UV irradiation treatments on the germination kinetics of alfalfa seeds are compared in **Figure 1A**. Seeds from the control group exhibit a median germination time of 26 hours and a 100% germination rate within the 60-hour observation period, indicating their nondormant status and robust physiological viability. The curves related to UV-A and UV-B treatments do not show significant difference with the control group while seeds exposed to a UV-C treatment present a median germination time of 24 hours, a modest improvement of 2 hours. Interestingly, the seeds treated with CP present a median germination time as low as 18 hours, meaning that the germination starts 8 hours earlier than in the control group. By 36 hours, nearly all seed groups, regardless of treatment, have achieved full germination, indicating the positive impact of CP on germination kinetics. In addition, the healthy and uniform phenotypes observed across all treatment groups (**Figure 1B**) suggest that neither CP nor UV treatments adversely affected embryo viability.

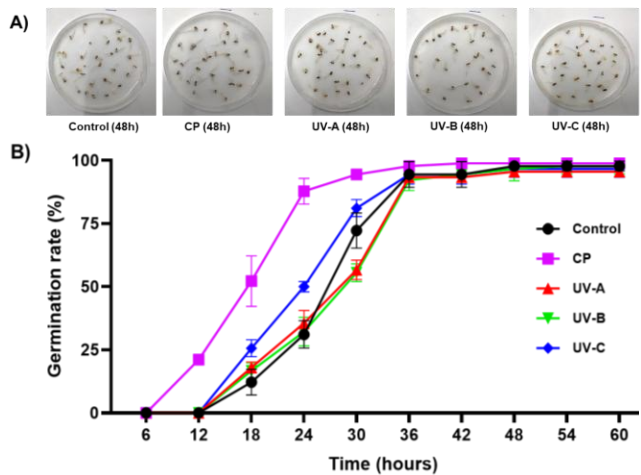

Figure 1. Germination kinetics of alfalfa seeds under different treatments and phenotypic comparison. (A) Germination kinetics of alfalfa seeds over a 60-hour period, comparing the effects of cold plasma (CP) and UV radiation (UV-A, UV-B, and UV-C) treatments with the control group. (B) Seed germination phenotypes at 48 hours after sowing for control, CP, UV-A, UV-B, and UV-C treated seeds. Each horizontal line indicates a 1 cm scale.

These later results, obtained by CP of ambient air, are aligned with those obtained on alfalfa seeds exposed to low-pressure plasma processes, whether in the works of Jinkui *et al.* (2018) who utilized an RF discharge of air-helium gas mixture applied at a few tens of Watts for 15 seconds or those of Tang *et al.* (2016) who utilized a neon discharge applied at 20 W for 20 seconds.

Plant responses to UV radiation depend on the species, exposure duration, and the specific UV bands. For instance, Peykarestan and Seify (2012) showed that seed germination rate of red bean was inversely correlated to the irradiation wavelength. Similarly, it has been demonstrated that germination rate and germination speed diminished as UV-B exposure time augmented from 5 minutes to 60 minutes (Ozel et al., 2021). However, Mariz-Ponte et al. (2018) reported beneficial effects of moderate UV-A supplementation on tomato seeds, accelerating and synchronizing germination. In conclusion, while low-intensity, short-duration (15 min) UV-C irradiation can accelerate germination, its effects are less pronounced compared to the stimulation achieved by the CP treatment.

## III.2. Variation in EL, seed imbibition and surface hydrophilicity

The assessment of EL from seed tissues is commonly used as an indicator of membrane integrity. As shown in **Figure 2A**, this parameter is increased in 72 hours, whether for untreated or treated seeds. Among the treatments, CP induced the highest EL at all time points (24 hours, 48 hours, and 72 hours), followed by UV-A treatment. In contrast, the seeds treated by UV-B and UV-C showed the lowest EL, suggesting better membrane stability. Interestingly, WU exhibited the opposite trend: seeds treated with CP showed the lowest WU (**Figure 2B**) while its surface coating exhibited the lower WCA values (**Figure 2C**). In turn, UV-B and UV-C treatments led to the highest WU and the lowest EL (**Figure 2A** and **2B**), hence illustrating better hydration and membrane stability in these treatments.

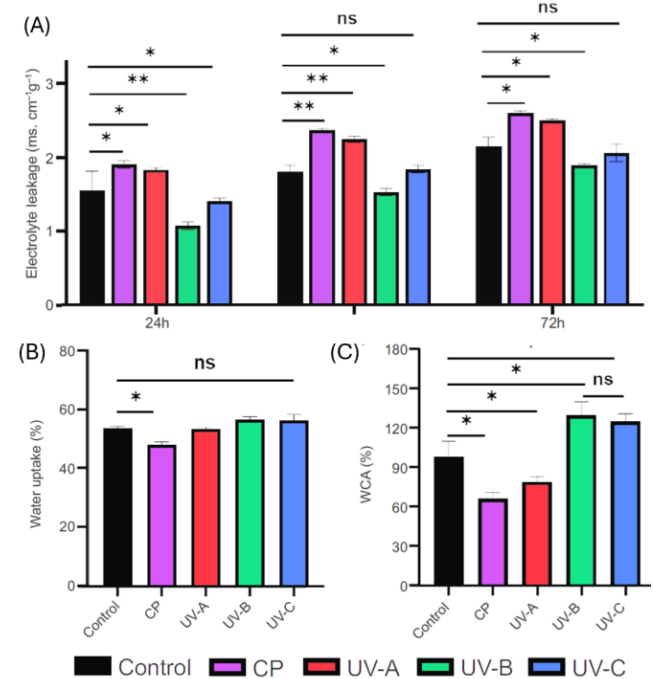

*Figure 2. Effects of different treatments on electrolyte leakage, water uptake, and water contact angle (WCA) of alfalfa seeds. (A) Electrolyte leakage of seeds and seedlings after 24 hours, 48 hours, and 72 hours (B); water uptake of seeds after 24 hours; and (C) WCA of seeds. CP: cold plasma; UV: Ultraviolet radiations A, B, and C. Asterisks indicate statistically significant differences between treatments (\*: P ≤ 0.05; \*\*: P ≤ 0.01; \*\*\*\*: P ≤ 0.001) according to the Duncan's test.*

## III.3. Growth of sprouts and biomass production

The observed interplay between EL, WU, and WCA highlights the critical role of membrane integrity and hydration in seed performance. Since these factors directly influence the availability of water and nutrients necessary for early seedling development, it is essential to examine how they affect subsequent biomass accumulation and sprout growth. As shown in **Figure 3A** and **3B**, the fresh weight (SFW) and dry weight (SDW) of sprouts were significantly higher in seeds treated with CP and UV-A compared to the control, UV-B, and UV-C treatments. Specifically, after 6 days, CP-treated seeds produced 25.87 mg/sprout fresh weight and 1.45 mg/seedling of dry weight, while UV-A-treated seeds yielded 23.77 mg/sprout fresh weight and 1.28 mg/sprout dry weight.

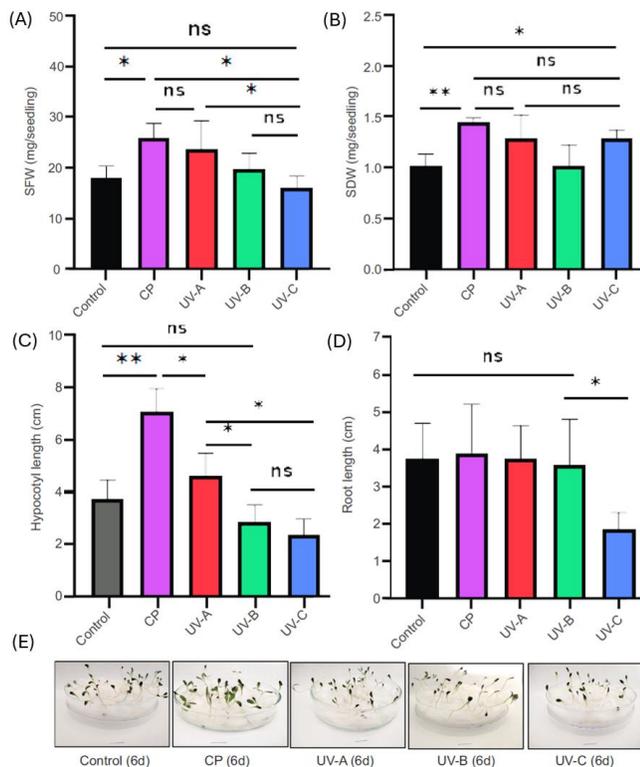

*Figure 3. Variation of sprout fresh weight (A), sprout dry weight (B), hypocotyl length (C), and root length (D) after 6 days from sowing variation according to treatments. Sprouts phenotypes after 6 days from sowing for each treatment (E). C: control; CP: cold plasma; UV: Ultraviolet radiations A, B, and C. The horizontal line indicates the 1 cm scale. Asterisks indicate significant difference between seeds treatments (\*: P ≤ 0.05; \*\*: P ≤ 0.01; \*\*\*\*: P ≤ 0.001) according to the Duncan's test.*

In terms of sprout growth, CP treatment also promoted significant hypocotyl elongation (**Figure 3C**), indicating enhanced development. In contrast, UV-C treatment notably reduced both HL and RL (**Figure 3D**), likely because of the damaging effects of UV-C's high energy and short wavelength on DNA and proteins, as reported by Beck *et al.* (2017). No significant difference in RL was observed between the other treatments. It is also interesting to observe a direct correlation between SFW (**Figure 3A**) and HL (**Figure 3C**), with these parameters showing exactly the same trends from one treatment to another.

Given the observed improvements in sprout fresh and dry yields, CP technology appears to be a promising method to improve the production of alfalfa sprouts. These stimulatory effects are similar to those obtained on other plant species, including soybean (Ling *et al.*, 2014), barley (Benabderrahim *et al.*, 2024), and wheat and oat (Sera *et al.*, 2010). The overall stimulatory effects of CP on seedling growth may be linked to changes in physiological processes like photosynthesis, which is further explored by measuring Chl content in different categories of sprouts.

## III.4. Variation of photosynthetic pigments

In higher plants, Chl a and Chl b are the most abundant pigments in the light-harvesting antenna system (Tang *et al.*, 2023). They play a central role in absorbing light and transferring it to the photosynthetic reaction centers. These pigments are essential for optimizing photosynthetic efficiency. **Table 1** summarizes the levels of Chl a and Chl b and total Chls in alfalfa sprouts under different treatments, while **Figure 4A** illustrates the variation of carotenoids' pigments. Significant differences in the accumulation of pigments were observed according to the treatment categories. The highest concentrations of the photosynthetic pigment were accumulated in sprouts from CP-treated seeds with total Chl reaching 16.18 mg/g FW and carotenoids reaching 5.09 mg/g FW, markedly higher than those in the control and UV-treated groups. In comparison, sprouts from the control seeds contained only 3.42 mg/g FW of total Chl and 2.01 mg/g FW of carotenoids. Among the UV treatments, UV-C showed a moderate increase in total Chl (5.82 mg/g FW), while UV-A and UV-B treatments resulted in lower pigment accumulation. The lowest values were recorded for UV-A- and UV-B-treated seeds, with 2.01 mg/g FW and 2.37 mg/g FW total Chl, respectively.

*Table 1. Photosynthetic pigments (mg/g FW) of alfalfa sprouts obtained from the different seed treatments (cold plasma and UV radiations).*

|         | Chl a        | Chl b       | Total Chl    |
|---------|--------------|-------------|--------------|
| **Control** | 1.84±0.02c  | 1.58±0.11b  | 3.42±0.08c   |
| **CP**      | 10.91±0.13a | 5.26±0.04a  | 16.18±0.09a  |
| **UV-A**    | 1.71±0.19c  | 0.28±0.01c  | 2.01±0.11d   |
| **UV-B**    | 2.07±0.31c  | 0.29±0.02c  | 2.37±0.46d   |
| **UV-C**    | 4.26±0.44b  | 1.56±0.42b  | 5.82±0.86b   |

Tang *et al.* (2023) reported that the key factors affecting the net photosynthetic rate were Chl content (especially Chl b content) and stomatal conductance, thereby explaining why increasing photosynthetic efficiency was critical for increasing crop yield (Ort *et al.*, 2015). Previous studies have consistently reported an increase in the accumulation of photosynthetic pigment in seedlings grown from CP-treated seeds. For example, Motrescu *et al.* (2024) demonstrated that CP treatment enhances Chl content in alfalfa sprouts, though the effect diminishes with prolonged treatment durations. Similarly, Fiutak *et al.* (2024) found that pulsed light stress applied during sprout cultivation can further enhance the accumulation of photosynthetic pigments and proteins. The exposure duration applied in the present study appears to be optimal for stimulating pigment production, thereby supporting biomass accumulation.

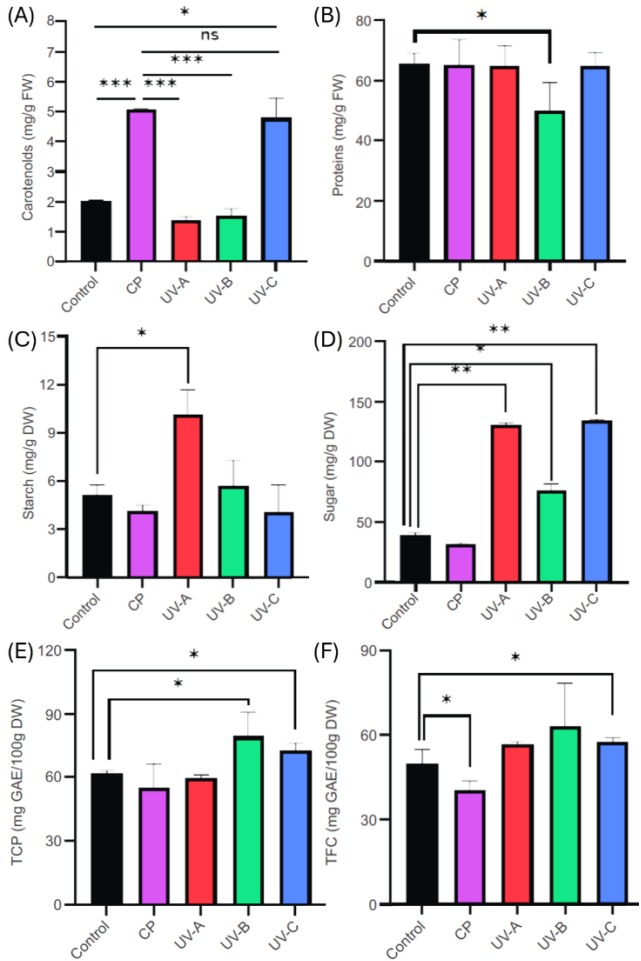

*Figure 4. Effects of different treatments on carotenoids (A), proteins (B), total of polyphenols (C), total flavonoids contents (D), starch content (E), and sugar content (F) in alfalfa sprouts. C: control; CP: cold plasma; UV: Ultraviolet radiations A, B, and C. Asterisks indicate significant difference between seeds treatments (\*: $P \leq 0.05$; \*\*: $P \leq 0.01$; \*\*\*\*: $P \leq 0.001$) according to the Duncan's test.*

CP-induced enhancement of Chl content likely contributes to improved photosynthetic performance and, consequently, higher biomass production (Mildaziene *et al.*, 2022). UV-C treatment, although less effective than CP, also increased total Chl content compared to the control, which correlates with a modest improvement in seedling biomass. These results suggest that CP enhances sprout biomass by boosting photosynthesis, the primary determinant of plant productivity.

### III.5. Effect of seed treatments on proteins and carbohydrate composition of sprouts

The protein content in alfalfa sprouts did not significantly vary between treatments (**Figure 4B**), except for UV-B, which induced a slight reduction in protein content (50 mg/100g DW) compared to the control (65.55 mg/100g DW). In a previous study, it has been confirmed that the penetration of UV-B radiation into the plant cells is limited, affecting surface or near-surface areas in plant cells (Liu *et al.*, 2013). Probably, this slight reduction of protein could be because of the RONS generated by UV-B seed treatment of alfalfa. In some cases, the intracellular components like DNA and protein may be damaged by generated ROS and RNS in the air atmosphere (Niedźwiedź, *et al.*, 2022). In contrast, Foroughbakhch *et al.* (2019) reported that UV-C radiation could induce a decrease in protein synthesis and the alteration of their structures. Even if few studies have investigated the effects of UV-B on protein content in plants, the reduction obtained in **Figure 4B** may result from several factors such as (i) inhibition of stress-related proteins, (ii) reduced enzyme activity, and (iii) altered protein expression.

To further investigate whether the CP and UV treatments influence the carbohydrates composition of sprouts, starch contents and soluble sugars (sucrose and β-glucan) were determined in the different seedlings. The obtained results are reported in **Figure 4C** and **4D**. Except for the UV-A treatment, no significant variation in starch contents was observed between treatments and control. Similarly, the total starch content remained unchanged after UV-B treatments in winter wheat (Yao *et al.*, 2014). The sprouts obtained from the untreated seeds accumulated an amount of starch of 5.12 mg/g DW. However, the UV-A treated seeds produced alfalfa sprouts with more starch (9.23 mg/g DW).

In terms of soluble sugar contents (**Figure 4D**), when comparing all treatments, the UV-A and UV-C treatments induced the highest sugar contents with mean values of 130.87 and 134.59 mg/g DW, respectively. These were followed by UV-B at 76.72 mg/g DW, while CP treatment showed the lowest soluble sugar concentration in sprouts (32.02 mg/g DW). Similar results have been obtained in tomatoes, where 1-hour UV-A treatment led to increased sugar levels (Mariz-Ponte *et al.*, 2021). The surprising rise in sugar content in alfalfa sprouts obtained from seeds treated with UV radiation can be explained through several hypotheses related to physiological and biochemical processes: (i) the accumulation of sugars and other osmolytes as part of the plant's stress response, (ii) the upregulation of genes expression related to carbohydrate metabolism, and (iii) starch breakdown induced by UV radiation.

Previous studies have also reported that dark treatment of alfalfa sprouts stimulates an increase in soluble sugar content (Sun *et al.*, 2024), further supporting the hypothesis that stress-related responses, such as UV exposure, could enhance sugar accumulation in sprouts.

### III.6. Effects of treatments on TPC and TFC

Polyphenolic compounds are a distinct class of naturally occurring molecules enriched with multiple phenolic functions. **Figure 4E** and 4F show the variations of TPC and TFC, respectively, under different treatments. Compared to the control, TPC remained unaffected in sprouts from CP-treated seeds but increased significantly in those treated with UV-B and UV-C. The TFC followed a similar trend, with higher flavonoids content observed in UV-treated sprouts. These results are consistent with the work of Kotilainen *et al.* (2008) who reported that exposure of gray alder (*Alnus incana*) and white birch (*Betula pubescens*) leaves to solar UV radiation alters flavonoid composition. This suggests that UV radiation, including UV-A, UV-B, and UV-C, stimulates the production of polyphenols in alfalfa sprouts, as part of a defense mechanism against UV-induced oxidative stress. As a result, seedlings may have higher levels of polyphenols following the treatment with UV irradiation. In addition, the variation in TFC may be attributed to the differential regulation of genes involved in flavonoid biosynthesis under UV radiation, as noted in legumes by Yan (2020). Consequently, the increase in polyphenols because of UV irradiation could impact the nutritional and medicinal properties of the alfalfa sprouts. Higher polyphenol content may increase the antioxidant activities of the seedlings, but it may also affect their growth and development. In this study, a 5-minute UV treatment appeared beneficial for polyphenol accumulation (**Figures 4E** and **3F**). However, this enhancement came with trade-offs, as evidenced by growth inhibition (**Figure 3C** and **3D**). These results highlight a balance between improving phytochemical content and maintaining growth, emphasizing the need for tailored UV treatments to maximize benefits while minimizing adverse effects.

### III.7. Comparative analysis of CP and UV radiation for enhancing alfalfa sprouts

To evaluate the effects of CP and UV treatments and uncover the mechanisms underlying their promotion of seed germination and sprout growth, we summarized the physiological and biochemical impacts of these treatments on alfalfa sprouts in **Figure 5**.

CP treatment demonstrated pronounced positive effects on both the physiological and biochemical properties of alfalfa sprouts. CP-treated seeds exhibited significantly enhanced photosynthetic pigment levels, with total Chl content (16.18 mg/g FW) and carotenoids (5.09 mg/g FW) being the highest among all treatments (**Table 1**, **Figure 4A**). These results indicate that CP-induced reactive oxygen and nitrogen species (RONS) activate metabolic pathways involved in Chl and carotenoids biosynthesis. The oxidative stress from CP likely stimulates these biosynthetic routes, enhancing light-harvesting efficiency and photosynthesis, which supports superior biomass accumulation (**Figure 3A** and **3B**). In addition, CP significantly reduced WCA, indicating improved seed surface hydrophilicity and potential for WU, though WU was paradoxically lower, likely because of cuticular modifications limiting water absorption. Furthermore, CP induced a shift in metabolic priorities, as evidenced by reduced soluble sugar content (32.02 mg/g DW, **Figure 4D**), likely redirected toward energy-demanding growth processes rather than storage. Despite these metabolic changes, protein levels remained comparable to the control, underscoring the ability of CP to stimulate growth without impairing primary metabolic functions.

UV-A treatment elicited moderate yet distinct metabolic responses in alfalfa sprouts. It significantly increased both starch (9.23 mg/g DW) and soluble sugar contents (130.87 mg/g DW, **Figure 4C** and **4D**), suggesting enhanced carbohydrate metabolism. This accumulation likely reflects an adaptive response to UV-A-induced mild oxidative stress, which promotes energy storage and osmolyte production to mitigate stress. UV-A also improved surface wettability by reducing WCA, potentially facilitating nutrient absorption and seedling development. However, unlike CP, UV-A had minimal effects on photosynthetic pigments, with total Chl content (2.01 mg/g FW) significantly lower than the control (**Table 1**). This suggests that UV-A primarily influences carbohydrate pathways rather than photosynthetic efficiency. The observed modest improvements in seedling biomass (**Figure 3A** and **3B**) and hypocotyl elongation (**Figure 3C**) align with these metabolic shifts, highlighting UV-A's role in balancing growth and stress mitigation.

UV-B treatment had a more complex impact, inducing moderate stress responses. While total Chl content (2.37 mg/g FW) remained low, UV-B significantly increased soluble sugar content (76.72 mg/g DW, **Figure 4D**), reflecting a shift toward osmoprotective mechanisms. These sugars likely play a dual role as energy sources and osmolytes, enhancing stress tolerance. In addition, UV-B treatment increased flavonoid synthesis (**Figure 4E** and **4F**), which contributes to photoprotection by scavenging reactive oxygen species (ROS) and absorbing harmful UV wavelengths. The slight reduction in protein content (**Figure 4B**) could result from oxidative damage to proteins or inhibition of protein synthesis because of stress. Despite these biochemical adaptations, UV-B-treated sprouts showed limited improvement in biomass accumulation (**Figure 3A** and **3B**), indicating that the metabolic cost of stress response may outweigh growth benefits.

UV-C treatment induced the most severe oxidative stress among the UV treatments, reflected in the substantial accumulation of soluble sugars (134.59 mg/g DW, **Figure 4D**). This high sugar content suggests an intense stress response aimed at mitigating oxidative damage by stabilizing cellular structures and maintaining osmotic balance. In addition, UV-C-treated sprouts exhibited moderate increases in total Chl (5.82 mg/g FW) and carotenoids (4.83 mg/g FW, **Table 1**, **Figure 4A**), which enhanced photoprotection and ROS scavenging. However, UV-C significantly inhibited seedling growth, as evidenced by reduced fresh and dry weights (**Figure 3A** and **3B**) and shorter HL and RL (**Figure 3C** and **3D**). This inhibition of growth likely results from the high-energy radiation of UV-C causing extensive DNA damage and impairing cell division and elongation. Interestingly, despite these growth limitations, protein content remained unaffected (**Figure 4B**), suggesting that UV-C-induced stress does not compromise protein biosynthesis but shifts metabolic priorities toward defense rather than growth.

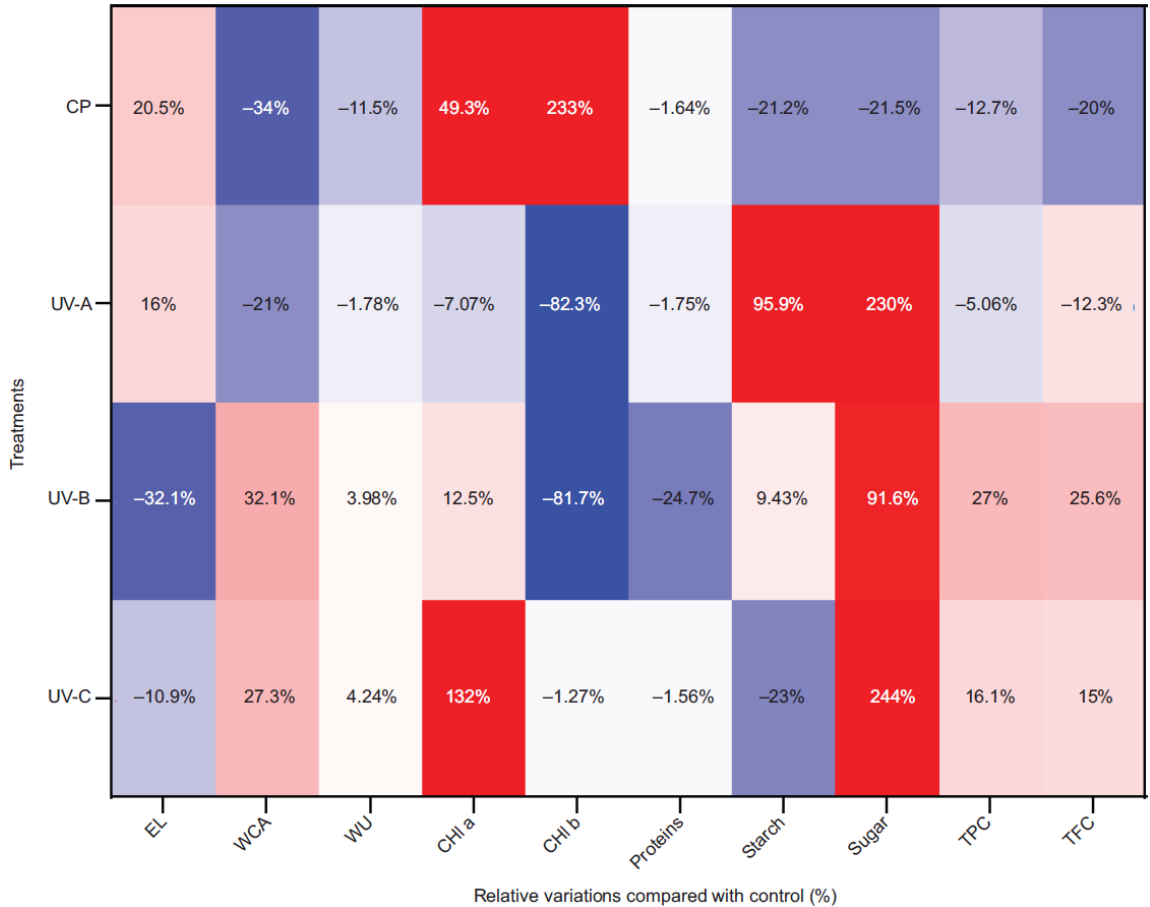

*Figure 5. Heatmap summarizing relative changes in physiological and biochemical traits of alfalfa sprouts treated with cold plasma and UV radiation compared to control.*

## IV. Conclusion

This study has demonstrated the potential of CP exposure and UV radiation as physical technologies for improving alfalfa seed germination, seedling growth, and phytochemical composition. CP treatment significantly improved germination kinetics, leading to faster and more uniform germination, with seeds germinating 8 hours and 6 hours earlier than untreated and UV-treated seeds, respectively. CP treatment also promoted superior seedling growth, resulting in the highest fresh and dry biomass, and a notable increase in Chl and carotenoid contents, indicating its ability to enhance photosynthetic efficiency. These results suggest that CP is a promising method for improving the production of alfalfa sprouts. On the other hand, UV radiation, particularly UV-C, showed potential for increasing specific phytochemical compounds, such as phenolics and flavonoids, which are known for their antioxidant properties. While UV treatments were less effective in improving overall growth, they did improve the nutritional profile of alfalfa sprouts by increasing soluble sugar, phenolic and flavonoid content, with UV-A and UV-C treatments showing the highest sugar accumulation. This highlights the potential of UV radiation to enhance the nutritional and medicinal value of crops. This work opens several avenues of research. Optimizing treatment parameters for CP and UV treatments could further enhance their effectiveness. In addition, studying the combination of these treatments could lead to synergies that optimize both growth and phytochemical content. Extending these technologies to other crops and assessing their long-term impact on crop quality and seed storage potential could provide valuable insights into their wider applicability. Finally, field trials and economic analyses would be essential to determine the feasibility of large-scale implementation of these technologies in farming practices.

## V. Data Availability

Data will be made available on request.

## VI. Author Contributions

MAB was responsible for writing the original draft, performing experiments, formal analysis, and investigation. HH did phytochemical analysis, writing, revising, and statistical analysis. WE was involved in writing, editing, and validation. TD was concerned with plasma treatments, writing, and editing and validation.

## VII. Conflicts of Interest

The authors confirm the absence of any conflicts of interest and disclose no significant financial support that may have biased the study findings.

## VIII. Funding

This work was supported and funded by the Deanship of Scientific Research at Imam Mohammad Ibn Saud Islamic University (IMSIU) (grant number IMSIU-DDRSP2502).